# The Negative-Energy Sea

Simon Saunders

## 1 Introduction

The Dirac hole theory was developed in response to a growing crisis over the Dirac theory of the electron. It predicted the existence of antiparticles in relativistic quantum theory; the antiparticle came into existence as a 'hole' in a sea of negative-energy particles. In 1972 Heisenberg said, 'I think that this discovery of antimatter was perhaps the biggest jump of all the big jumps in physics in our century'. He was speaking of the phenomenology of pair creation and annihilation processes, the basic mechanisms of relativistic dynamics, but the conceptual basis, the *concept* of antimatter, has a corresponding importance.

If this concept was initially tied to the negative-energy sea, that is not the case any longer. The negative-energy sea remains a widespread heuristic device to introduce antimatter, and a review of the hole theory is still given in most elementary textbooks on relativistic quantum theory, but nowadays no one would claim that the negative-energy sea actually exists; it is no longer taken as a literal description of the vacuum. How is it, then, that the hole theory can be dispensed with? What takes its place? We know that in some sense antiparticle states are related to negative-energy states; relativity leads to antimatter because the constraint $E^2 - p^2 c^2 = m^2 c^4$ is satisfied by negative energies as well as positive energies. The question is, in what precise sense, if not in the sense of the Dirac hole theory?

One of the most widespread heuristics, due to Feynman and Stueckelberg, identifies antiparticles with negative-energy particles moving backwards in time. Antimatter arises just because it is possible for positive-energy particles to scatter backwards in time with negative energies, emitting energy in the process (pair annihilation). This heuristic finds a natural expression in the path-integral approach to quantum theory. However, this approach marks a decisive break with the canonical theory (in particular, with the exact Hilbert space theory). A good reason to be interested in the canonical formulation of the concept of antimatter is to understand better the relationship between the relativistic and the non-relativistic theory. For these reasons I shall omit discussion of the Feynman-Stueckelberg interpretation.

---

[§] [Note added March 2024. Some typos and infelicities have been corrected. I am publishing online for the sake of wider accessibility, and because part of the content is historical, and may be of enduring interest (§2-6 are purely expository). I also believe there is more to be learned in philosophy of QFT from the 'two complex structures' perspective of §7 (although I no longer think germane to the measurement problem, contra the suggestion in fn.43). I add two references: one, by me, on locality and the bosonic case ('Locality, complex numbers, and relativistic quantum theory', *Proc. Phil. Sci. Assoc. Vol.1* (1992), p.365-380), and one by David Wallace, on the relation of $U(1)$) symmetry to other symmetries ('QFT, antimatter, and symmetry', *Stud. Hist. Phil. Mod. Phys.* **41**: 209-222).]

In what follows I shall first sketch the history of the Dirac theory; afterwards I shall concentrate on a more recent development in the canonical framework which, I believe, throws new light on both the hole theory and the field theory which replaced it. This development has its origins in the Segal theory of quantization; what is characteristic of this approach is that complex numbers are built into the classical solution manifold in a geometric way, and this manifold is then identified with the 1-particle Hilbert space. In this way the negative-energy sea is encoded into the mathematical description of the antiparticle states. Something like this was already achieved in the mid-1930s, but mediated by the fields; the Segal theory makes explicit a Hilbert space analogue. Because my concern is to explore the interpretation of quantum electrodynamics within the canonical framework, I shall consider only the linear quantum electrodynamics. Dirac himself was led to the hole theory purely on the basis of the linear equation; in the linear case we already see all the important features of relativistic quantum theory.

In elementary quantum mechanics the vacuum is very simple; it is the quantum analogue of the Newtonian vacuum. In the vacuum not only are there no particles, but there is no theory. There is no Hilbert space, there is no time evolution, one cannot write down equations for this vacuum. The vacuum concept (as distinct from the concepts of space and time) can be described only informally. We have the same situation in classical particle mechanics. But in quantum field theory (also in continuum mechanics and classical field theory), the vacuum is modelled in the mathematics. One might say that in these theories 'what exists' becomes a dynamical quantity, for which non-existence takes on the value zero. (As one value among others, the vacuum must be modelled in the mathematics.) The idea of 'vacuum' is relativized to the observable content of the theory, be it states of a medium, excitations of a field, or particle number. In quantum electrodynamics, despite the field aspect, the vacuum is defined not as the zero-valued fields (there is no state in which all the fields have eigenvalue zero), still less as a zero-valued wave function (which is not even a state), but rather in terms of the absence of any particles. The canonical vacuum is the state of emptiness of particles.

It might seem that this concept of vacuum is essentially unique, and almost as simple as in the elementary theory. Every particle observable has the value zero with probability one.[1] Nevertheless, there are self-adjoint operators for which this is not the case, for example certain combinations of the quantum fields. From the point of view of these operators, the vacuum is not at all trivial. Properties of the vacuum picked out in this way may still be interpreted in particulate terms (almost entirely, in perturbation theory), and there is a direct connection with the picture of the quantum field as a collection of harmonic oscillators (zero-point energy); but it seems to me that a more immediate problem is to understand why such operators arise in particle mechanics in the first place. In particular, one wants to understand how in the Dirac theory even well defined particle observables are *required* to have vacuum expectation values that are non-zero (and in fact infinite).

There is a more general problem. As I have indicated, the Dirac vacuum brings in its wake the concepts of antimatter and pair creation and annihilation processes. These transform the quantum theory into an edifice of remarkable phenomenological expressiveness and real mathematical complexity. The mathematical framework of non-relativistic quantum field theory was reasonably well understood by the late 1920s;[2] almost a century later, the simplest of (non-trivial) relativistic theories still resists any comparable elucidation. My objective is this: to characterize better those features of relativistic theory that are responsible for this pathology.

In the historical review of Sections 2-6 we observe shifts in the theoretical perspective in due chronological sequence; however, the framework is throughout tied to fermionic theories. The characterization proposed in Sections 7 and 8 fits naturally into this framework but in fact applies equally to fermionic and bosonic theories.

The latter results are restricted; they apply only to global kinematic observables. This theory is, however, exact. It must be born in mind that the conventional theory can be made rigorous only in the kinematic limit; the Fourier analysis is available only for the free field, and in its absence, one has no precise particle interpretation.

---

[1] More precisely, every self-adjoint operator that can be defined as the canonical second quantization of a particle operator has eigenvalue zero in the vacuum state. This is true of the non-relativistic theory; it is also true of the theory developed in Section 7, without recourse to normal-ordering.

[2] I have in mind the proof of equivalence of the interacting Galilean field theory with a many-particle ensemble due to Jordan and Klein (1927); see also Tomonaga (1962). The detailed analysis of Fock space methods came somewhat later (Fock 1932; Cook 1953); these and later developments in the mathematical theory of quantum fields are irrelevant to the present discussion, because one can always restrict the field theory to a finite-particle subspace of the Fock space. (This is not possible in the relativistic case.) For applications of non-Fock representations, see Saunders (1988).

## 2 The Origins of the Hole Theory

In 1928 Dirac wrote down a wave equation, which is Lorentz- covariant and first-order in the time:

$$[i\hbar\gamma^\mu\partial_\mu - mc]\psi(\boldsymbol{x}, t) = 0.$$

For an external c-number field with potential $A_\mu(\boldsymbol{x}, t)$, one then has, for a particle of charge $-e$,

$$\left[\gamma^\mu\left(i\hbar\partial_\mu - \frac{e}{c}A_\mu(\boldsymbol{x}, t)\right) - mc\right]\psi(\boldsymbol{x}, t) = 0.$$

In these equations $\gamma^0, \gamma^1, \gamma^2, \gamma^3$ are 4 x 4 complex matrices, and $\psi$ is a 4-component complex-valued function on space-time. The $\gamma$ matrices provide a representation of the Clifford algebra which is unique up to isomorphism. They satisfy $\gamma^\mu\gamma^\nu + \gamma^\nu\gamma^\mu = [\gamma^\mu, \gamma^\nu]_+ = 2g^{\mu\nu}$ . The function $\psi$ is usually called a Dirac spinor, or bispinor. Dirac was led to this equation because he was looking for a first-order analogue of the Klein -Gordon equation, namely:

$$(\Box + m^2c^2/\hbar^2)\phi(\boldsymbol{x}, t) = 0$$

where $\phi$ is a complex scalar function and $\partial_\mu\partial^\mu = \Box$ is the d'Alembertian operator. For an external electromagnetic potential $A_\mu$, this becomes

$$\left[\left(i\hbar\partial_\mu - \frac{e}{c}A_\mu(\boldsymbol{x}, t)\right)\left(i\hbar\partial^\mu - \frac{e}{c}A^\mu(\boldsymbol{x}, t)\right) - m^2c^2\right]\phi(\boldsymbol{x}, t) = 0.$$

The Dirac equation is a linearized square root of this equation; that is, there are linear combinations $P, P'$ of $\partial, A$ and $m$ such that $P\psi = 0$, and such that $P'P\psi = 0$ is the Klein-Gordon equation. For this to be possible, $P, P'$ must contain matrices, and correspondingly $\psi$ must be a many-component object. Dirac found that it is not possible to linearize the square-root equation with two-dimensional matrices; the minimum dimension is four, and then one has the Clifford algebra. This follows from the requirement

$$\left(-i\hbar\gamma^\mu\partial_\mu - mc\right)(i\hbar\gamma^\nu\partial_\nu - mc) = \hbar^2\Box + m^2c^2 .$$

Dirac wanted a first-order equation because he thought that only then could one find a probability interpretation, and define a transformation theory, as in the non-relativistic quantum mechanics. In retrospect, it is clear that he sought a Schrödinger equation,[3] which must indeed be first-order in time; but Dirac also demanded covariance, which is to ask too much. The result is a wave equation which, used as a

---

[3] I use the term to mean the infinitesimal form of the unitary time evolution with positive generator on the Hilbert space of states.

Schrödinger equation, leads to a theory that is much more than a sum of its parts.

Initially Dirac had only a fragmented formalism; defining the free Hamiltonian by analogy to Schrödinger theory, he obtained the operator($i = 1, 2, 3$)

$$H_D = -i\hbar\gamma^0\gamma^i\partial_i + \gamma^0 mc^2.$$

Likewise, he considered the quantity

$$\int \sum \overline{\psi(\boldsymbol{x})}\,\psi(\boldsymbol{x})d^3x$$

the probability density. (The summation is over the spinor components.) Unfortunately, although this density is positive definite, the spectrum of the free Hamiltonian is not; formally, there exist functions $w\exp[-i(Et - \boldsymbol{p}\cdot\boldsymbol{x})/\hbar],\, w\in\mathbb{C}^4,$ which satisfy the wave equation for both signs of $E$.

Initially this formalism yielded some striking results: it predicted the correct $g$-factor for the electron and the Sommerfeld equation for the spectral lines of the hydrogen atom. For these reasons, the negative-energy difficulty did not lead to the abandoning of the theory. It was acknowledged from the beginning; Dirac (1928) suggested that the negative-energy solutions might correspond to positive-charge particles, and that they could be rejected on this basis.

A few months later he conceded that they could not simply be excluded from the theory, because in the presence of interactions there might be transitions from positive- to negative-energy states and these could not be eliminated by fiat. The theory was then to be thought of as an approximation. But increasingly, it became clear that the difficulty could not be contained or restricted to any non-trivial dynamical regime. Heisenberg, one of the first to perceive the extent of the departure from the principles of quantum mechanics, went so far as to remark: 'the saddest chapter of modern physics is and remains the Dirac theory'.[4] Much later he was to say: 'up till that time I had the impression that in quantum theory we had come back into the harbor, into the port. Dirac's paper threw us out into the open sea again' (Heisenberg 1963). At this stage one couldn't modify the mathematics too much because there seemed to be too much truth contained in the theory.

There were two further developments that made the difficulty of negative-energy states that much more acute. One was the demonstration, due to Oskar Klein (1929), that even for time-independent potentials there may be no solution of the Dirac equation with only positive-frequency parts (the 'Klein paradox'). The other was the discovery, made independently by Igor Tamm (1930) and Ivor Waller (1930), that the negative-frequency parts played an essential role in the classical limit of the Klein-Nisjima scattering formula; that is, in order to get the Thomson formula, it was necessary in perturbation theory to sum over intermediate negative-frequency states.

---

[4] Letter to Pauli, 31 July 1928 (Heisenberg 1928).

These states had to be taken seriously. Dirac saw from the beginning that they had to correspond to particles of opposite charge. In the Klein-Gordon case, which also admits negative-energy solutions, there had been suggestions to the same effect.[5]

But a consistent interpretation proved elusive; classically, a negative-energy negative-charge particle behaves in an external electric field just as a positive-energy positive-charge particle; if the electromagnetic potentials are reversed, it actually behaves in an *identical* way to its positive-energy partner. Its space-time world-line would be identical. But surely, this is a peculiarity of conservative systems; intuitively it seems clear that a negative-energy particle will have to emit energy as it speeds up, and that is unphysical. Dirac could not just posit that the negative-frequency states are positive-charge positive-energy states.[6]

If one thinks about negative energy, one has to work in terms of the absence of positive energy; and if one also thinks about positive charge, one might be led to think of it as the absence of negative charge. From that, it is a short step to the idea that a particle of positive charge and positive energy might correspond to the *absence* of a particle of negative charge and negative energy.

This notion, that the absence of a particle is a physical thing, and has a dynamical role in the theory, had already been employed in quantum theory, and has precursors in classical physics.[7] The most familiar example is the Bohr theory, where one has electron transitions to orbits that are not closed, which do not have their full complement of electrons. Dirac cited internal conversion, where an inner electron is ejected by absorption of X-rays, and remarked that this absence of an electron is described by a wave function and plays much the same role as a physical particle.

The difference is this: there is nothing analogous to the almost-filled Bohr orbitals; there is nothing with respect to which this absence may be defined. It was here that Dirac made a truly revolutionary hypothesis. The negative-energy particles indeed exist, but they exist everywhere and in such abundance that in general transitions to such states will be forbidden by the Pauli exclusion principle. At the same time, if there were an available negative-energy state, it would appear as an absence of negative energy and negative charge, and hence (relative to the background) as a particle of positive energy and positive charge. If there are very few missing negative-energy electrons, and if these transactions are the only empirical manifestation of the existence of the negative-energy sea, then we would scarcely be aware of its existence.

---

[5] See e.g. Fock's (1926) derivation where he used a proper time parameter, and the Klein (1926) derivation on 5-dimensional space-time.

[6] This reasoning is contained in Dirac (1929). Nevertheless, the identification is made out in Section 7; the problem posed by Dirac is eliminated by use of Segal's methods.

[7] For example, discussing an analysis due to J. J. Thomson of the magnetic field associated with a charged moving conductor, G. F. FitzGerald (1881) showed that the displacement currents set up in the ether by the time-varying electric field could not be circuital. As an example, he considered a charged parallel-plate capacitor; if one plate approaches the other, the electric field is 'annihilated' by the plate, the electric displacement is therefore destroyed, and there must exist a corresponding displacement current. (This current evidently has non-zero divergence; FitzGerald then showed that the total current, including that arising from the motion of the charged plate, is circuital.)

The assumption of the negative-energy sea is extravagant, even by the standards of the physics of our day. One assumes that each finite volume of space has infinite charge and infinite energy, to make conceptual sense of the theory. In an illuminating remark, Wightman was later to comment:

> It is difficult for one who, like me, learned quantum electrodynamics in the mid 1940s to assess fairly the impact of Dirac's proposal. I have the impression that many in the profession were thunderstruck at the audacity of his ideas. This impression was received partly from listening to the old-timers talking about quantum-electrodynamics a decade-and-a-half after the creation of hole theory; they still seemed shell-shocked. (Wightman 1972: 99)

Familiarity breeds tolerance; one suspects that for later generations it is not so much that the negative-energy sea is considered a fiction, but that no categorical basis seems to exist by which mathematical artifice may be distinguished from the reality.

To understand the significance of the Dirac vacuum, one has to explore the mathematical background of quantum theory at a deeper level. To understand the immediate context in which Dirac worked, one has to understand the second quantization process, and his theory of the equivalence of the quantized electromagnetic field with a many-boson system. The second quantization will play an important role in all that follows, so I shall start with this theory.

### 3 Canonical Second Quantization

Starting from a canonical 1-particle theory, with a Hilbert space $\mathfrak{h}$, one defines creation and annihilation operators as maps between $n$- and $(n+1)$-particle spaces, which are constructed as symmetrized or anti-symmetrized tensor products of the 1-particle Hilbert space. The total Hilbert space must contain all these finite particle subspaces, so it is of the form

$$\mathcal{H} = \mathfrak{F}_S(\mathfrak{h}) := \sum_n S_n \otimes_{i=1}^n \mathfrak{h}_i$$

Here $S$ indicates the appropriate symmetrization and $S_n$ is a representation of the symmetrization operator for the permutation group of order $n$. (Later on we suppress the subscript $S$.) Each $\mathfrak{h}_i$ is a copy of $\mathfrak{h}$.

To make the connection with field theory, it is essential that the particles are identical; that is, the states that are built up by successive applications of the creation operator cannot contain information as to which particle is in which state. This being so, the set of occupation numbers, a string of integers $n_1 \, n_2 \, \ldots \, n_i \, \ldots$, is enough to parameterize these states, where each subscript $i$ determines a particular state $\phi_i$ of the 1-particle theory and each occupation number $n_i$ fixes the number of particles in that state. (We suppose the states $\phi_i$ form an orthonormal basis for $\mathfrak{h}$.) In terms of this parameterization, the action of the annihilation and creation operators is just

$$a(\phi_i)\colon |n_1 \ldots n_i \ldots\rangle \to (n)^{1/2} \, |n_1 \ldots (n_i - 1) \ldots\rangle \,.$$

$$a^*(\phi_i)\colon |n_1 \ldots n_i \ldots\rangle \to (n+1)^{1/2} \, |n_1 \ldots (n_i + 1) \ldots\rangle \,.$$

(The normalization constants are slightly different in the antisymmetric case.) These operators are adjoints of each other, as our notation suggests; as a result, if one is a linear map on the 1-particle space $\mathfrak{h}$, the other must be *antilinear;* that is, for any complex scalar $\lambda$

$$a^*(\lambda\phi) = \lambda a^*(\phi)$$

$$a(\lambda\phi) = \bar{\lambda} a(\phi) \,. \tag{1}$$

The antilinearity of the annihilation operator is so important that it is helpful to see why it holds in an intuitive way. For a state $\eta \in \mathfrak{I}_S(\mathfrak{h})$ of the form $f_1 \otimes f_2 \otimes f_3 \otimes \cdots \otimes f_n \oplus$ permutations, and arbitrary $f \in \mathfrak{h}$, we have

$$a(f)\eta = (n)^{1/2}\langle f, f_1\rangle f_2 \otimes f_3 \otimes \cdots \otimes f_n \oplus \text{permutations} \tag{2}$$

The antilinearity of the annihilation operator is therefore a consequence of the antilinearity of $\langle \,.\,,.\,\rangle$ in its first entry, the hermitean inner product on $\mathfrak{h}$.

Using these operators, one can write down the operator on an n-particle Hilbert space which corresponds to a 1-particle operator $A$ on $\mathfrak{h}$, such that this operator makes no reference to particle identity, namely

$$\overbrace{\underbrace{A \otimes \| \otimes \cdots \otimes \|}_{\text{n-fold tensor product}} \oplus \| \otimes A \otimes \cdots \otimes \| \oplus \cdots \oplus \| \otimes \cdots \otimes \| \otimes A}^{\text{n-fold tensor sum}} \,. \tag{3}$$

For an arbitrary orthonormal basis $\{\phi_i\}$, an equivalent definition is

$$\sum_{i,j} a^*(\phi_j)\langle \phi_j, A\phi_i\rangle a(\phi_i) \,.$$

The important point about this operator is that it duplicates the action of

(3) on any *n*-particle subspace (i.e. whatever the value of *n*); therefore this expression, but not (3), can be used in a theory in which the particle number is specified along with the state, that is as part of the specification of the initial conditions. This is a first step towards generalizing the theory to deal with dynamical situations in which the particle number is variable.

The operator $\sum_{ij} a^*(\phi_i)\langle\phi_j, A\phi_i\rangle a(\phi_i)$ is called the *second quantization* of the 1-particle operator *A*; it is usually denoted dΓ(*A*). dΓ is a structure-preserving map ('functor') independent of the orthonormal basis use in its definition. In particular, self-adjointness and positivity are preserved by dΓ. If *A* generates the unitary group *U*, then we *define* the group generated by dΓ(*A*) as the second quantization of *U*, which we denote Γ(*U*). It follows that

$$\Gamma(U)d\Gamma(A)\Gamma(U)^{-1} = d\Gamma(UAU^{-1}) \tag{4}$$

which provides an important class of unitary evolutions on $\mathfrak{F}_S(\mathfrak{h})$ (i.e. those determined by unitary 1-particle evolutions on $\mathfrak{h}$).

The creation and annihilation operators can also be used to construct the total number operator; the quantity $\sum_i a^*(\phi_i)\,a(\phi_i)$ applied to an *n*-particle state returns that state unaltered, except that it is multiplied by the constant *n*; likewise, the operator $a^*(\phi_i)a(\phi_i)$ is the number operator for the state $\phi_i$. Note that the total number operator is the second quantization of the identity; i.e.,

$$d\Gamma(\mathbb{I}) = \sum_{i,j} a^*(\phi_j)\langle\phi_j, \mathbb{I}\phi_i\rangle a(\phi_i) \tag{5}$$

The number operator for the state $\phi_i$ is the second quantization of the projection operator on to the subspace spanned by the state $\phi_i$.

The transformation theory can be applied to these quantities in a rigorous way; formally, one often uses the improper position and momentum eigenfunctions also.

dΓ(*A*) has a simple interpretation. Applied to any many-particle state, it gives the appropriate action of the 1-particle operator on each particle in an ensemble. On states of the form $\eta$, the c-number $\langle\phi_i, A\phi_j\rangle$ under the summation is multiplied into each c-number $\langle\phi_j, f_1\rangle$ left as residue of the annihilation of the 1-particle state $f_1$ (cf. (2)), and the state $\phi_i$ is returned to the state vector $\eta$ in its place; since we sum over all $i, j$, we evaluate the total transition amplitude for each particle under the influence of *A*. This construction also works for 2-, 3-,or *n*-particle operators; for example, a second quantized 2-particle operator dΓ(*B*) is

$$d\Gamma(B) = \sum_{ijkl} a^*(\phi_i)a^*(\phi_j) \ll \phi_i\phi_j, B\phi_k\phi_l \gg a(\phi_k)a(\phi_l)$$

(where I have written $\phi_i\phi_j$ for the symmetrized 2-particle state and $\ll ., . \gg$

for the induced inner product on the 2-particle subspace). Figure 1 provides a graphic illustration of the action of this operator.

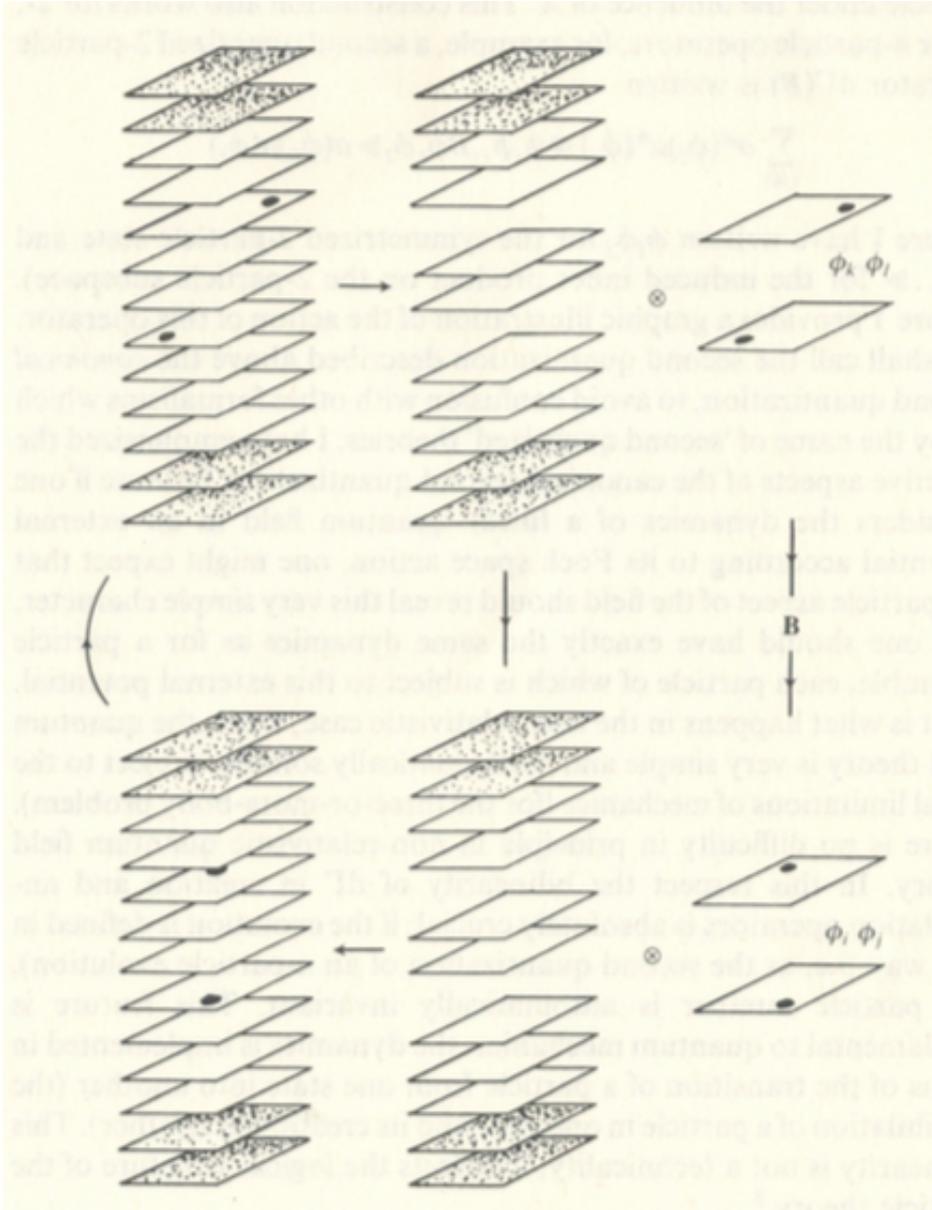

FIG.1  The action of the operator

$$\sum_{ijkl} a^*(\phi_i) a^* \left(\phi_j\right) \ll \phi_i \phi_j, B\phi_k \phi_l \gg a(\phi_k) a(\phi_l)$$

I shall call the second quantization described above the *canonical* second quantization, to avoid confusion with other formalisms which go by the name of 'second quantized' theories. I have emphasized the intuitive aspects of the canonical second quantization, because if one considers the dynamics of a linear quantum field in an external potential according to its Fock space action, one might expect that the particle aspect of the field should reveal this very simple character, and one should have exactly the same dynamics as for a particle ensemble, each particle of which is subject to this external potential. That is what happens in the non-relativistic case; there, the quantum field theory is very simple and mathematically soluble, subject to the usual limitations of mechanics (for the three-or-more-body problem). There is no difficulty in principle in non-relativistic quantum field theory. In this respect the bilinearity of $d\Gamma$ in creation and annihilation operators is absolutely crucial; if the evolution is defined in this way (i.e. as the second quantization of a $n$-particle evolution), the particle number is automatically invariant. This feature is fundamental to quantum mechanics; the dynamics is implemented in terms of the transition of a particle from one state into another (the annihilation of a particle in one state and its creation in another). This bilinearity is not a technicality; it reflects the logical structure of the particle theory.[8]

However, this is not at all the situation in the relativistic case, even for a linear field coupled to an external potential. The problem arises entirely from the existence of negative-energy states; the simplest interactions connect positive- and negative-energy states. Nowadays it is usually said that there is no 1-particle relativistic quantum mechanics, so of course the field theory is not the canonical second quantization of any 1-particle theory. Rather, using perturbation theory to read back from the field theory, one learns that what happens on the particle level is pair creation and annihilation, which cannot be understood in terms of particle dynamics (excluding Feynman-Stueckelberg methods). •

This point of view arose in the mid-1930s, but Dirac had already considered a similar problem in 1927. He attempted to extend the canonical particle framework so as to describe *individual* absorption and emission processes. This is worth a closer look.

## 4 The First Dirac Vacuum

One might say that Dirac sought to provide a precise particle interpretation of the emission and absorption of light quanta by atoms,[9] an aspect of Einstein's great 'heuristic hypothesis', which after more than twenty years was at last to find full mathematical expression. Dirac had formulated a quantum electrodynamics which led to such processes by applying his theory of action-angle variables, developed in his relativistic q-number (matrix) mechanics of the Compton effect in 1926, transferred to the radiation field in place of the mechanical atom.[10] At the same time, he had discovered that similar techniques applied to the non-relativistic (linear) Schrödinger equation yielded a formalism equivalent to the quantum mechanics of a many-particle boson system (the canonical second quantization which I have just described). Dirac tried to bring the two into correspondence: obviously, the difficulty is the mechanical description of the creation and annihilation of photons.

[8] There is also a connection with group theory, which leads to the Bargmann mass superselection rule. Essentially, the mass arises not as a Casimir invariant, but in the choice of the central extension of the Galilean group; see e.g. Sudarshan and Mukunda (1974)

[9] There are many routes to Dirac's (1927) development of quantum field theory (wave-particle duality, quantum electrodynamics, the correspondence principle, the Einstein theory of $A$ and $B$ coefficients, the Kramers-Heisenberg dispersion theory, the Compton effect); see Saunders (1989 Part 1) for further commentary.

The problem is simple. Because Dirac wanted to describe the back-reaction on the field, the potential $V$ is no longer an external perturbation, and must be considered a dynamical entity in its own right, with its own degrees of freedom. It becomes, in fact, a quantum field. However, it immediately appears - in support of Einstein's original conceptions - that the Hamiltonian derived from this potential contains an isolated creation (or annihilation) operator.[11] How to relate this theory to the canonical theory? There one can only obtain operators of the form $d\Gamma(A)$; necessarily, these are bilinear in creation and annihilation operators, which is only to say that net particle creation and annihilation processes cannot be described in the mechanical theory.

We have another way of understanding this difficulty if we look at Dirac's proof of the equivalence between a $q$-number linear Schrödinger equation and a boson ensemble. It is implicit here that the $c$-number Schrödinger equation for each boson is formally identical to the $q$-number equation. When one tries to do the same thing for the radiation field coupled to charge, the interaction Lagrangian contains a term linear in the vector potential; hence the field equations are inhomogeneous and no longer linear.

Consider for example the electrostatic case; there is an interaction density of the form $V(\boldsymbol{x}, t)\rho(\boldsymbol{x}, t)$, with $V$ the electrostatic potential and $\rho$ the charge density. (In the non-relativistic theory, the latter is bilinear in creation and annihilation operators: no net particle creation or annihilation here.) If one looks at the field equations, the $q$-number version of what should be the 1-particle Schrödinger equation is obviously nothing of the kind. Because the Lagrangian contains the term $\rho V$, the field equations contain the inhomogeneous term $\rho$, one has Poisson's equation $\nabla V = -e\rho$, which cannot be understood as a $q$-number version of a Schrödinger equation. The equivalence between Schrödinger and field equations exploited by Dirac has disappeared.

---

[10] In his theory of the Compton effect he had used the action-angle operators to describe transitions of the atom, not of the field. The action-angle operators provided the algebraic structure of the creation and annihilation operators; see Dirac (*1926a, 1926b*).

[11] According to Dirac, the reality of the potential required its expansion as the sum of creation and annihilation operators for the (positive-frequency) light-quanta. Dirac used a non-relativistic description (for photons!). The negative-frequency difficulty does not arise.

We are faced with a difficulty which is perhaps more acute than that of the negative-energy states. (Dirac began with the hard problems.) In order to get out of this difficulty and preserve a relationship between the canonical second quantized theory and the quantized field theory, Dirac supposed that photon number is *also* conserved; the apparent annihilation of a photon is in reality the process in which a photon of frequency $v$ makes a transition to a photon of frequency zero. That is, there are many (infinitely many) zero-frequency photons present all the time, and individual creation and annihilation processes become photon transitions to and from this new vacuum. In this way Dirac described the quantized electromagnetic field as a particle theory; this was his resolution of the wave-particle duality as he traced it to Einstein.[12] There is no precise mathematical basis to this theory, but we have the first indication that the dynamics can be changed dramatically by modification of the vacuum. This was already clear in 1927.

## 5 The Negative-Energy Sea

Whereas one might accept that a vacuum filled with zero-frequency photons is still a nothingness, because a zero-frequency photon is just like a vacuum state of a classical field, the situation is different when the vacuum is full of massive, charged electrons. What of the gravitational properties of this vacuum? How does this vacuum respond to electromagnetic fields? Even if there are no holes present, there must be other empirical consequences of this idea. What can it mean to have infinite charge and mass in finite volume?

It was a bold and radical and quite outrageous suggestion. Pauli, even after the discovery of the positron, was absolutely against it. As he later put it, 'success seems to have been on the side of Dirac rather than of logic'. We have seen Heisenberg's reaction;[13] Bohr also was sceptical. It

---

[12] Dirac observed that the 1-particle 'analogue' of photon creation and annihilation (transition to and from the zero-frequency state) could not be written 'as an algebraic function of canonical variables'. In this sense Dirac did not achieve a complete mechanical description of the interaction (contrast with the hole theory). On this point, von Neumann was later to take an anti-realist stand: 'It is difficult to find a direct, clear-cut, interpretation of the interaction energy…nevertheless, we can accept this with the interpretation that each model-description is only an approximation, while the exact content of the theory is furnished solely by the expression for the Hamiltonian operator' (von Neumann 1932: 282-3; the 'model-description' is the particle theory, the Hamiltonian is derived from the field).

[13] In early 1934 Heisenberg wrote: 'I regard the Dirac theory…as learned trash which

is well known that Anderson, the discoverer of the positron, was indifferent to the Dirac theory (Anderson and Anderson 1983). The theory was exotic and speculative; prior to the discovery of the positron, only a handful of experts were concerned with it.

But the hole theory was successful; its heuristic power was immense. If there is a hole present, then a positive-energy electron may make a transition to this state with the emission of energy. This would appear as the annihilation of both the electron and the hole. If a negative-energy electron absorbed energy, so that its total energy became positive, it would leave behind it a hole in the negative-energy sea, and this would appear as the creation of a positive-energy electron together with a hole. The inferences follow effortlessly. Phenomena of this kind were soon observed.

The sequence of events went something like this. To begin with, physicists were not then disposed to predict the existence of new kinds of particles. There was only one sub-atomic positively charged particle known, and that was the proton. Weyl had already suggested that the negative-frequency states describe the proton; Dirac rejected their identification with protons, for reasons already summarized, but he suggested that these might appear as the holes in the negative- energy sea. The mass difference, he surmised, may be accounted for by the interaction among all the negative-energy electrons. This was in November 1929; by March of the following year, he (and, independ- ently, Oppenheimer) calculated the cross-section for electron-proton decay. Even given the ambiguity introduced by the electron-proton mass difference, the result was much too large to be consistent with the stability of ordinary matter. The same conclusion was reached by Tamm one month later; by the end of the year Weyl had gone into print retracting his earlier suggestion. In May 1931 Dirac predicted the existence of the positron. It was observed by Anderson that summer.[14]

The way was open to evaluate scattering cross-sections for a number of new phenomena (all to be experimentally observed). Between 1930 and 1935 the following processes were considered (the calculations used $c$-number potentials for the radiation):

$e^+ + e^- \rightarrow \gamma$ (Dirac, Oppenheimer, Tamm)

$\gamma + \gamma \rightarrow e^+ + e^-$ (Breit, Wheeler)

$e^+ + e^- \rightarrow \gamma \rightarrow e^+ + e^-$ (Bhabba)

$\gamma + \gamma \rightarrow e^+ + e^- \rightarrow \gamma + \gamma$ (Halpern)

The fundamental fact is that, by redefining the ground state of the electron

---

no one can take seriously' (1934a ). Three months later he was able to eliminate it from the formalism (see below).

[14] This sequence of events is well documented; see e.g. Bromberg (1976) and Pais (1986) for further details and references.

theory, the standard ideas of 1-particle quantum theory[15] immediately lead to a phenomenology typical of a many-particle theory. The idea of pair creation and annihilation had actually been around for some time; some progress had been made in studying the equilibrium properties of such processes using semi-classical arguments. But with the hole theory, pair creation and annihilation became 1-particle processes; there is the interpretation, and there is the mathematical formalism. Immediately one could calculate cross- sections and apply the perturbation theory to deduce the existence of more complex processes that proceed by virtual states (such as the Bhabba and Halpern scattering). It is often said that the Dirac hole theory transforms the 1-particle theory into a many-particle theory; we see that it also works the other way: prima facie many-particle processes involving pair creation and annihilation can be treated using the formalism of the 1-particle theory.

In the processes that we have considered the negative-energy sea plays a purely passive role, in restricting the number of negative-energy states available for such transitions. In other processes the sea is more active: its response to an external field should be just like a dielectric. The negative-energy electrons will be polarized and an induced polarization field will be set up. This is the *vacuum polarization*, first investigated by Dirac in 1934, and here for the first time the full intricacies of the hole theory were encountered. To deal with them Dirac used a variant of the Hartree self-consistent field method. This step leads naturally to the calculation of the effective charge that will produce the 'net' field, that is the external field together with the polarization field of the vacuum; in other words, he was led to the idea of *charge renormalization.* This is the first time that the notion of renormalization entered quantum physics.[16]

# 6 The Standard Formalism

With these developments, the Dirac hole theory became part of the basic vocabulary of physics. Every practicing high-energy physicist today knows the theory; it has an apparently enduring heuristic role. But the negative-energy sea is no longer considered a literal description of reality. What, then, became of the theory, and how do we do without it today?

---

[15] Here, by '1-particle system' I mean a particle that may be found in positive- or negative-frequency states. The term 'particle-antiparticle system' is unsatisfactory, because it suggests that one has a particle pair; rather than introduce this or an even more cumbersome terminology, I shall leave it to the context to distinguish the 1-particle system in the present sense from a 1-particle system defined over positive- frequency states alone. No confusion will result.

[16] In other situations, e.g. the interacting non-relativistic field considered by Jordan and Klein (1927: 761-2), the quantum theory effectively *removes* a renormalization problem of the classical theory; the self-energy of the classical field theory disappears through normal-ordering. (This is the only application of normal-ordering to the non- relativistic theory. See Saunders (1989: Part 1).

There seem to be two answers to this question. The first is that we have learnt to dispense with this heuristic and to rely on the mathematics unaided. The second is that we have a new theory, formally similar, based on the Feynman-Stueckelberg heuristic and path-integral methods.

This new theory I will have to place on one side. It is the first response that concerns us; it is based on what used to be called the second-quantized hole theory, or the Jordan-Wigner formulation, but now it is called the Dirac field theory. I shall call it the *standard* formalism. The fundamental step was to reformulate the hole theory as a second-quantized theory (in a rather imprecise sense). This step was taken by a number of people in 1934--by Fermi, in connection with the theory of $\beta$-decay; by Heisenberg, in order to eliminate the negative-energy sea and the asymmetry between positive and negative charge; and by Fury and Oppenheimer, in their systematic reconstruction of the Dirac hole theory. It was reinforced by the new impetus to field theory provided by Pauli and Weiskopf, Yukawa, and the rapid growth of meson physics from the mid-1930s. I shall not discuss these developments; I shall only present the second-quantized version, more or less as did Heisenberg (1934b).

To begin with, consider the canonical second-quantization. If for our orthonormal basis we use instead the 'improper' momentum eigenfunctions, we are led to the `operator` $\beta_r(p)$ annihilating an electron (of positive or negative energy) of four momentum $p$. The subscript $r$ picks out one of the two spin eigenstates with respect to a selected component of spin.[17] Taking the Fourier transform, extended over both halves of the mass shell,[18] we obtain the point field:

---

[17] There is an unfortunate complication here in connecting the standard formalism to the canonical theory. The details will not be relevant to what follows, but in parenthesis let me say this: the plane-wave expansion was first written down for the 1-particle solutions of the Dirac equation. Therefore the bispinor appears explicitly. On 'second-quantizing', the expansion coefficient $b$, was made into an annihilation operator. In the canonical framework, it is simpler to work with annihilation operators of the form $a(\phi)$, with $\phi$ a 1-particle state (and if possible avoid a parametrization of the space of states). In that case $\phi$ includes the bispinor. (This is what we shall do in Section 7.) If one treats instead the quantities $b_r(p)$ as annihilation operators, then they must act on the Fock space over the $s = \frac{1}{2}$ spinor representations constructed by Wigner (1939) (and not over the solution space of the Dirac equation). This feature of the standard formalism is perhaps not a technicality when it comes to perturbation theory; in kinematics, however, it is well understood.

[18] For the time being we can consider the Fourier transform an application of the transformation theory in the canonical theory; viewed in this way, we must expand over a complete set of states, therefore over both positive- and negative-energy states.

$$\psi(x) = \int_{E>0} \sum_r w_r(p) b_r(p) \mathrm{e}^{-ip\cdot x/\hbar} \mathrm{d}\mu^+$$

$$+ \int_{E<0} \sum_r w_r(p) b_r(p) \mathrm{e}^{-ip\cdot x/\hbar} \mathrm{d}\mu^- .$$

Here the summation is over the two linearly independent spin states, and $\mu^\pm$ is the invariant measure on the mass shells (which includes constants in $\hbar$ and $\pi$). It is usual to rewrite this by letting $r$ run over four values, one pair for each sign of the energy, i.e. for positive-energy solutions with $r = 1,2$

$$w_r(\boldsymbol{p}) = w_r\big((\boldsymbol{p}^2 + m^2c^2)^{1/2}, \mathbf{p}\big)$$

and for negative-energy solutions $w_3, w_4$:

$$w_{r+2}(\boldsymbol{p}) = w_r\big[-(\boldsymbol{p}^2 + m^2c^2)^{1/2}, \boldsymbol{p}\big]$$

(and similarly for $b$). One can then carry out the integral over the energy $p_0$, obtaining the familiar form:

$$\psi(x) = (2\pi\hbar)^{-3/2} \int_{\mathrm{p}^3} \left[ \sum_{r=1,2} b_r(\boldsymbol{p}) w_r(\boldsymbol{p}) \mathrm{e}^{-ip\cdot x/\hbar} \right.$$

$$\left. + \sum_{r=3,4} b_r(-\boldsymbol{p}) w_r(-\boldsymbol{p}) \mathrm{e}^{-ip\cdot x/\hbar} \right] \frac{\mathrm{d}^3 p}{\sqrt{2}p_0}$$

(here $p_0 = +(\boldsymbol{p}^2 + m^2c^2)^{1/2}$; the normalization is $\bar{w}_r w_s = 2p_0/c\delta_{rs}$).

We now consider the canonical second quantized operators. By formal application of the transformation theory (using 'improper' position eigenstates[19]), any operator $A(\boldsymbol{x})$ which is local in the 1-particle theory (a multiplicative function or involving only finite derivatives in configuration space coordinates) can be written in the form $\int \sum \psi^*(\boldsymbol{x}) \psi(\boldsymbol{x}) \mathrm{d}^3 x$. Applied to the 1-particle Hamiltonian $H$, one obtains, after some manipulation:

$$\mathrm{d}\Gamma(H) = \int \sum_{r=1,2} p_0 \left[ N_r^+(\boldsymbol{p}) - N_r^-(-\boldsymbol{p}) \right] \mathrm{d}^3 p/p_0 ,$$

where we define the number operators:

$$N_r^+(\boldsymbol{p}) = b_r^*(\boldsymbol{p}) b_r(\boldsymbol{p}) \text{ (for positive-frequency states)}$$

---

[19] These are doubly improper, since they have little to do with particle position. This part of my treatment is undoubtedly clumsy, and should be replaced by a more fundamental treatment of the Fourier transform. However, I wish to avoid a detour into the representation theory of abelian groups; interested readers are referred to Mackey (1963) for a general perspective.

$$N_r^-(\boldsymbol{p}) = b_{r+2}^*(\boldsymbol{p})b_{r+2}(\boldsymbol{p}) \text{ (for negative-frequency states)}.$$

As in the 1-particle theory, the canonical second quantized energy is indefinite. The total charge is the second quantization of $-e\mathbb{I}$ and one verifies that

$$\mathrm{d}\Gamma(-e\mathbb{I}) = -e\int\sum_{r=1,2}[N_r^+(\boldsymbol{p}) - N_r^-(-\boldsymbol{p})]\mathrm{d}^3\,p/cp_0\,.$$

This is negative definite, as we expect (since the charge of electrons, whether positive- or negative-frequency, is negative).

So much for the *canonical* second quantization. Despite the formal manipulations, everything can be made rigorous and put into the canonical framework. But if we now consider the action of the negative-energy creation and annihilation operators on the Dirac vacuum, the negative-energy sea, clearly the annihilation operator will create a hole (positron) and the creation operator will annihilate a hole, or will give the value zero if there is no hole present. Accordingly, let us change our notation; we shall replace $b_3(-\boldsymbol{p})$ by $d_1^*(\boldsymbol{p})$ and $b_4(-\boldsymbol{p})$ by $d_2^*(\boldsymbol{p})$. (The change in sign in $\boldsymbol{p}$ is for convenience; using the symbol $d$ rather than $b$ eliminates the need for the index values 3 and 4, and the * indicates whether we are dealing with a creation or annihilation operator with respect to the positrons.) Similarly, $b_{r+2}^*(-\boldsymbol{p})$ is replaced $d_r(\boldsymbol{p})$; the anticommutation relationships obeyed by these operators are unchanged by these substitutions. (This would not be true if they obeyed commutation relationships.)

The effect of these substitutions is that when we evaluate the quantities $d\Gamma(H), d\Gamma(-e\mathbb{I}))$, we obtain the quantities as above except that now $N_r^-(-\boldsymbol{p}) = d_r(\boldsymbol{p})d_r^*(\boldsymbol{p})$; *the order of the creation and annihilation operators is reversed.* That being so, this term *cannot be interpreted as the positron number operator.*

This can easily be remedied; we use the anticommutation relationships (ACRs) to write these quantities (the original number operators for negative-energy electrons) in terms of number operators for positrons. The latter are given by the quantities $N_r^-(\boldsymbol{p}) = d_r^*(\boldsymbol{p})d_r(\boldsymbol{p})$; in this way we obtain $d_r(\boldsymbol{p})d_r^*(\boldsymbol{p}) = -N_r^-(\boldsymbol{p})$ + positive infinite constant. The expressions for the total energy and charge *(E* and *Q)* become:

$$E = \int\sum_{r=1,2}p_0[N_r^+(\boldsymbol{p}) + N_r^-(\boldsymbol{p})]\mathrm{d}^3\,p/p_0 - \text{positive inf. const.}$$

$$Q = -e\int\sum_{r=1,2}[N_r^+(\boldsymbol{p}) - N_r^-(\boldsymbol{p})]\mathrm{d}^3\,p/cp_0 - \text{positive inf. const.}.$$

The change in sign in these quantities is crucial. The infinite constants correspond, in the hole theory, to the infinite negative energy and negative charge of the negative-energy sea. The remaining contribution to the energy (charge) is now positive definite (indefinite). But as yet the spectrum of these operators is unchanged, nor could it be changed merely by a change in notation and use of the ACRs.

In quantum field theory it is standard practice to subtract such infinite constants (zero-point subtractions) produced by the reordering; the reordering followed by the setting to zero of all c-numbers is called *normal-ordering.* By this 'standard practice', however, *the spectrum of the operator is changed.* The subtraction makes the energy into a positive operator, and the spectrum of the charge operator is no longer negative definite. Mathematically, therefore, the zero-point subtraction is far from trivial. By this same practice, we also find that we have a new basis for the theory: *if one now writes down the momentum expansion for the quantum field $\psi(x)$ with the 'correct' interpretation of the observables, one need make no reference whatsoever to the Dirac hole theory.* That is, if from the word go we write the quantum field as

$$\psi(\boldsymbol{x}, t) = \int_{E>0} \sum_{r=1,2} w_r(p) b_r(p) e^{-ip \cdot x/\hbar} \mathrm{d}\mu^+$$

$$+ \int_{E<0} \sum_{r=1,2} w_{r+2}(p) d_r^*(-p) \mathrm{e}^{-ip \cdot x/\hbar} \mathrm{d}\mu^-$$

and normal-order the physical observables, we need never bother with the hole theory. The negative-energy sea has done its work once we have the 'correct' particle interpretation of the field, which is to say the plane-wave expansion above, and once we no longer demand a physical interpretation of the normal-ordering process.[20] In particular, using the Lagrangian theory and Feynman diagrams, we can develop a formal perturbation theory for interactions which lends itself readily enough to intuitive visualization. The relevant heuristic is that particles are created and destroyed in pairs, so as to preserve charge; these are not transition processes of a single particle involving negative- and positive-frequency states. Both intuitively and in the mathematics, we can no longer treat the resulting theory as the canonical second quantization of a 1-particle theory. The technique for making the energy positive (correct momentum space expansion + normal-ordering) does not seem to make any sense at the 1-particle level; a precise correspondence is lost. In this respect we have the same situation as in the Dirac hole theory (with the negative-energy sea as vacuum), but actually the situation is worsened; we have no physical basis for the rift with the canonical theory.

---

[20] This is to be considered a purely mathematical technique, which does not stand in need of justification. For an account along these lines, see Wightman (1972). In the theory of Section 7, the normal-ordering has a fundamental significance. Path-integral theory also places a more fundamental perspective on normal-ordering. (This is easiest to understand in Euclidean theory; see e.g. Simon 1974, sec. I.I).

Nowadays no one would regard the use of this Fourier expansion, or of the normal-ordering, as logically dependent on the Dirac hole theory; they are supposed to stand in their own right. At the same time, the entire theory can be regarded as a quantum field theory, and the link with the 1-particle theory becomes hopelessly tenuous. One looks upon the Dirac equation as a classical field equation, itself derived from a classical Lagrangian. The quantization of this theory is to yield the standard formalism (correctly interpreted), as above. There is, however, a connection between the antimatter fields and the negative-energy solutions. The latter contribute negative energy to the total field energy. (The use of anticommutators on quantization allows us to change the sign of this contribution, depending on whether we consider it a creation field or an annihilation field.) This part of the field $\psi$ is the antiparticle (creation) field. But the negative-energy solutions disappear from the Fock space description (there are no negative-energy states); there is a doubling-up of states, and their distinction is made at the $q$-number level. What were before calculations of transition amplitudes at the level of the states become analysis at the level of the fields; particularly, it is analysis on the c-number bispinors that occur in the plane-wave expansion.

It does not appear possible to understand antimatter at the level of the states. As a result, certain questions, such as the definition of states when one does not have a scattering situation, or the meaning of the Wigner negative-energy representations of the Lorentz group (which now appear to be excluded by fiat), cannot even be formulated. It is the canonical theory that imparts precision to these questions.

Let me pursue the question of the independence of the standard formalism from the Dirac hole theory. The problem is to justify the plane-wave expansion of the fields. It turns out that the necessary assumptions have a natural interpretation in field theory: the field must be a linear combination of creation and annihilation operators. This is implicit in some earlier discussions, but (so far as I know) there is no very clear statement prior to Weinberg (1964). One reason for this neglect is that the precise definition of the creation and annihilation operators - namely an explicit action on a concrete Fock space - was not and could not have been available prior to the late 1950s, because of the difficulty in relating the Dirac bispinors to the Wigner spinor representations. This problem (cf. fn. 17) requires the distinction between representations on Hilbert space and those on Hilbert space bundles;[21] the explicit bispinor $c$-numbers that occur in the plane-wave expansion should be understood as transformation matrices between the $C^4$ fibre sitting over the base space and the Wigner $C^2$-valued Hilbert space of spinors.

But neither was Weinberg concerned with the logical status of the plane-wave expansion. He was trying to demonstrate the independence of the *S*-matrix theory from Lagrangian methods. To this end he was driven to work from first principles. He found[22] that in kinematics, if one is to get covariant operators that commute (or anticommute) at spacelike separation, it is necessary to take linear combinations of creation and annihilation operators. Even then, it is not necessary that one have an annihilation operator for a particle state and a creation operator for an antiparticle state (they could be creation and annihilation operators for a single species of particle). However, in that case the field will (in general) be complex, but it will not transform in any simple way under (global) gauge transformations; it will not transform as $\psi \rightarrow \exp(i\theta)\psi$ with $\theta$ as c-number. The reason is that, if the field $\psi$ is a linear combination of creation and annihilation fields on the *same* Fock space, then if the annihilation field (say) transforms as $b \rightarrow \exp(i\theta)\psi$, the creation field must transform as $b^* \rightarrow \exp(-i\theta)b^*$ because it is the adjoint field. So the final upshot is that, if we want to have a covariant, causal, field that is complex, but transforms simply under gauge transformations, then we must introduce a new Fock space (the antiparticle space), and the linear combination of annihilation operators on the particle space and creation operators on the antiparticle space (together with its adjoint) is the only possible operator expansion.

---

[21] A crucial link that was first investigated in a physical application by Joos (1962).

Weinberg was happy to have isolated simple assumptions concerning the field, sufficient for the derivation of the plane-wave expansions; with these, the Feynman rules could be defined, and on the *S*-matrix philosophy no further appeal to a dynamical theory (such as the Lagrangian theory) was necessary. For our purposes, what is important is that these are just the properties of the field as required in Lagrangian theory; the field is covariant, causal, and gauge-covariant.

In this way the standard formalism can be considered logically independent of the Dirac hole theory. However, one must still motivate the normal-ordering process, and one finds that the antiparticle states have no relationship to the negative-energy 1-particle states. On the contrary, they are identical to the positive- energy states. This needs careful consideration.

The antiparticle states {the elements of the antiparticle Fock space) behave identically under Lorentz transformations, including space and time inversions, as the particle states. {In particular, they have positive energy.) One merely supposes that these states are *distinct,* so that $b_r(\boldsymbol{p}) \neq d_r(\boldsymbol{p})$. What makes them distinct? First and last, it is the action of the fields. It is the way the fields couple the two kinds of states that leads to the characteristic dynamics whereby pairs of particles (one in each class of states) are destroyed and created. What prevents single creation and annihilation processes, or pair processes of other kinds (each from the same class of states), is the requirement that the Hamiltonian be gauge-invariant.

---

[22] The summary that follows is a slight modification of Weinberg (1964), along the lines of Novozhilov (1975).

This is something new to the principles of elementary quantum mechanics, although its impact, in particular its implications for measurement theory and the transformation theory, is somewhat reduced under the rubric of charge superselection. It is often said that operators that connect states of different total charge do not exist. But the meaning of this statement is unclear. There is nothing comparable in the non-relativistic theory. The mass superselection rule has a different origin; the gauge invariance of the Hamiltonian is there a consequence of the fact that it is self-adjoint.

I have omitted mention of charge conjugation. One might think that the explicit definition of this operator will clarify the relationships between matter and antimatter on the one hand, and positive- and negative-frequency states on the other. It is even said that the introduction of charge conjugation restored the symmetry between positive and negative charge, and cleared the way to the elimination of the negative-energy sea. In fact, the charge conjugation adds little to our understanding, and involves additional complications.

In the Weinberg construction, this operator (denote $\mathfrak{C}$) is defined by the interchange of the $b$ operators with the $d$ operators, or simply by the interchange of the two Fock spaces (for particle and antiparticle). Since these are identical as function spaces, it is trivial that $\mathfrak{C}$ is unitary.

However, at the level of the fields it can be written in a way that is formally identical to the charge operator $\mathcal{C}$ in the 1-particle theory, which is antiunitary. The details are as follows. If one takes the adjoint (complex conjugation plus matrix transposition) of the Dirac equation in the presence of an external field

$$\left[\gamma^\mu\left(i\hbar\partial_\mu - \frac{e}{c}A_\mu(\boldsymbol{x},t)\right) - mc\right]\psi(\boldsymbol{x},t) = 0$$

one obtains the equation (superscript $t$ is matrix transpose)

$$\bar{\psi}^t(\boldsymbol{x},t)\left[\bar{\gamma}^{\mu t}\left(-i\hbar\partial_\mu - \frac{e}{c}A_\mu(\boldsymbol{x},t)\right) - mc\right] = 0\,.$$

From the defining properties of the $\gamma$ matrices, it follows that $\gamma^0\bar{\gamma}^{it}\gamma^0 = \gamma^i$, $i = 1,2,3$; so inserting a factor $\gamma^0\gamma^0$ between $\bar{\psi}^t$ and $\gamma^{\mu t}$, and operating from the right by $\gamma^0$, one obtains, on taking the matrix transpose,

$$\left[\gamma^{\mu t}\left(-i\hbar\partial_\mu - \frac{e}{c}A_\mu(\boldsymbol{x},t)\right) - mc\right]\gamma^{0t}\bar{\psi}(\boldsymbol{x},t) = 0\,.$$

This equation is not quite in the right form, because of the transposition of the $\gamma$ matrices. However, if there exists a matrix $C$ such that $C\gamma^{\mu t}C^{-1} = -\gamma^\mu$, we may insert a factor $C^{-1}C$ and operate from the left by $C$ to obtain:

$$\left[\gamma^{\mu}\left(i\hbar\partial_{\mu}+\frac{e}{c}A_{\mu}(\boldsymbol{x},t)\right)-mc\right]\psi^{c}(\boldsymbol{x},t)=0$$

where we have written $\psi^{c}=C\gamma^{0t}\bar{\psi}$ (the positron state). This is the Dirac equation for a particle of *positive* charge.

Actually, if $\psi$ is a positive-frequency state, then $\psi^{c}$ is a negative frequency state, so that to obtain positive-frequency solutions of the positive-charge Dirac equation we must take the charge conjugate of negative-frequency negative-charge solutions. The map $\mathcal{C}\colon\psi\to C\gamma^{0t}\bar{\psi}=\gamma^{c}$ is called the *1-particle charge conjugation operator*; a matrix $C$ with the defining property above exists and can be chosen unitary; because of the complex conjugation, however, is $\mathcal{C}$ antilinear (hence antiunitary).

However, it seems that one cannot consistently define charge conjugation within the 1-particle theory. The reason is that the total charge (and likewise the charge current density) do not change sign under the 1-particle charge conjugation. Since the charge operator is just $-e\mathbb{1}$, it is obvious that this operator is invariant under charge conjugation, contrary to physical requirements.

The relationship with the charge conjugation $\mathfrak{C}$ for the quantum field is as follows. The formal application of the operator $\mathcal{C}$ to the field takes one from the field to its adjoint, hence is equivalent to the interchange of $b$ and $d$, even though it defines transformations of the form $b\to b^{*},d\to d^{*}$. Since these transformations are antilinear, it seems we must have an antiunitary transformation (and an anti-automorphism with respect to the fields); but if $\mathcal{C}$ is automorphic and one normal-orders after its application, then one obtains the same transformation as $\mathfrak{C}$. The total charge now changes sign under charge conjugation, because (normal-ordered) it is no longer a multiple of the identity.

This is a curious situation; the charge conjugation appears at once antilinear at the level of the fields, yet unitary at the level of the Fock space. On the other hand, if the gauge transformation of the fields • is induced by that of the states, the particle and antiparticle states must stand in antilinear correspondence,[23] in contradiction to the unitarity of $\mathfrak{C}$.

The standard formalism presents puzzling features, and the relationship between the 1-particle and field-charge conjugation operators is one more example. The particle interpretation of the fields, from which follows all of the mathematical pathology of relativistic quantum theory, is secured if the fields are covariant, gauge-covariant, and causal, but these requirements make no sense at the 1-particle level. The distinction between matter and antimatter cannot be made out at the level of the states, and the negative-energy representations of the Lorentz group play no role in the theory. We are a long way from the clear-cut heuristics of the Dirac hole theory.[24]

---

[23] I.e., if the transformation $a(f)\to\exp(i\theta)a(f)$ arises from the transformation $f\to\exp(i\theta)f$, $f\in\mathfrak{H}$, and $f$ and $g$ are respectively particle and antiparticle states, then we must have $g\to\exp(-i\theta)g$.

Confronted with this situation, one feels a certain exasperation. Surely the antiparticle *field* is the negative-frequency *field* solution of the *field* equation, just as the negative-energy *state* is the solution of the *wave* equation. The connection between antimatter and negative energy should be direct and simple. In order to make it so, we must find a formulation of the canonical theory which directly relates q-number and c-number versions of the same equations. Fortunately, that has been worked out for us.

## 7 Quantization and Complex Numbers

I refer to the so-called *geometric quantization,* due to several workers, but above all to Irving Segal.[25] He was concerned specifically with the quantization of linear classical fields; in its more developed form (following Souriau 1966, and Kostant 1970) it provides a quantization process - and with it a representation theory - which generalizes the Dirac correspondence between commutation relationships and the Poisson bracket. This theory leads to a rigorous quantum theory of much more general systems, e.g. constrained systems on manifolds; however, we will make use only of the basic construction provided by Segal.

This construction is applicable in all cases where a rigorous Fock space representation has been established in quantum field theory. It can be considered a generalization of the canonical theory; its novel features disappear in the non-relativistic limit.

Suppose that we have a linear dynamical system and an associated phase space, that is a pair $V, \omega$ where $V$ is a real vector space[26] and $\omega$ a bilinear form, antisymmetric for bosons (the symplectic form) and symmetric for fermions; and let us introduce a canonical transformation $J$ such that $J^2 = -1$. With the aid of this, we can construct a *complex* vector space and a sesquilinear inner product, and can complete the vector space to obtain a Hilbert space[27] (denote $V_J$). The point of this construction is that *symplectic*

---

[24] These obscurities are eliminated in the theory that follows; I omit, however, a discussion of covariance and microcausality, which hinge on the analysis of locality (see Saunders 1989, sec. 3.4).

[25] See e.g. Segal (1964, 1967). For a review of the more general theory, see e.g. Woodhouse (1980).

[26] This does not mean that $V$ might not also be a complex vector space (i.e. that it is also complex-linear); the point is that we use only the real-linear structure.

[27] This procedure was foreshadowed in the work of Stueckelberg and his co-workers in the early 1960s; see e.g. Stueckelberg and Guenin (1962) and references therein.

*(bosons) or orthogonal (fermions) transformations on V which preserve J automatically become unitary transformations on $V_J$* . In particular, the Hamiltonian flow, the group of transformations on $V$ corresponding to the (classical) time evolution, becomes a weakly continuous group of unitary transformations on $V_J$ so long as it preserves $J$. The quantum mechanical Hilbert space $\mathcal{H}$ is then given as the space of *analytic[28]* functions on this space. In quantum field theory, $V$ is already a function space, the space of classical solutions to the classical field equations,[29] and $\mathcal{H}$ is naturally represented as the Fock space over $V_J$. The complexified phase space has a natural correspondence with the 1-particle subspace of $\mathcal{H}$. In the simplest case the dynamics is simply lifted from the classical Hamiltonian flow on $V$, so that we can regard the induced evolution as the canonical second quantization of a 1-particle evolution. In this way we can preserve a very close correspondence with the 1-particle theory (or, equivalently, with the c-number solutions to the field equations). Indeed, although we start from a field theory, the relationship of the field to the particle interpretation is the same as in the canonical second quantized theory.[30]

For interacting theories of physical interest (even linear theories) one cannot put this construction on any simple basis; in particular, there does not exist a canonical complex structure $J$ which is preserved under the time evolution. We see that the complex structure $J$, which tells us what we mean by complex numbers in the Hilbert space theory, also tells us what we mean by particle number (or more generally a particle interpretation) for a quantum field. The favourable case roughly coincides with the situation where the field is kinematic (it actually includes time-independent external couplings); otherwise we shall suppose that interactions lead to a change in $J$ and the particle interpretation is shifted; quanta have been created or destroyed.

In perturbation theory, too, one defines the asymptotic states (and, by the assumption of completeness, even the interacting states) in terms of the kinematic description of the quantum field. So this is a familiar limitation. What is the kinematic description? It is provided by the decomposition of the field into positive- and negative- frequency parts. Here the complex numbers that enter into the classical fields play a crucial role; for a given spacelike hypersurface $\ell$ and solution $\phi$, one finds functions $\phi^+$ and $\phi^-$ such that $\phi = \phi^+ + \phi^-$, and when $\ell$ is translated in time the complex phases of these functions on $\ell$ rotate in opposite directions.[31]

---

[28] Unitarity and analyticity are both defined with respect to $J$. This Hilbert space of analytic functions was first introduced by Fock (1928); see Bargmann (1961) for a systematic treatment. An analytic function has a power series expansion; when it is defined on a function space (as in classical field theory), the term in this expansion linear in vectors in $V$ is the 1-particle component of this state.

[29] More precisely, the space of Cauchy data for the field. For the Dirac field, this can be identified with square-integrable $\mathbb{C}^4$-valued functions on $\mathbb{R}^3$. The theory can be formulated in a covariant way, but we do not need this here.

[30] On this basis I shall at times speak of the field quantization as a canonical second quantization of a 1-particle theory. By this I mean no more than that from the field quantization one can read off the 1-particle theory, to which it is related by the functor $\Gamma$.

This complex structure, the $i$ that (may) appear in the field equations, I shall call the *natural complex structure.* These are the complex numbers that are usually used in Hilbert space theory. They are also the complex numbers that are used to define the gauge transformation properties of the fields (and to define the gauge-invariant objects in the theory, i.e. the observables). We recall that, if we switch this gauge transformation to the Hilbert space level, the particle and antiparticle states are rotated in opposite directions (cf. fn. 23). Multiplication by $i$ at the level of the fields is mirrored in the Hilbert space by multiplication by the imaginary unit (of the Hilbert space) on the particle states, and by minus the imaginary unit on the antiparticle states. We may conjecture that it is $J$ that determines what we mean by complex numbers in the quantum theory of a classical system, in particular by the particle interpretation of a quantum field, and that $J$ is related to the natural complex structure through the decomposition into positive- and negative-frequency parts. That is just what happens; $J$ is given by multiplication by $i$ for a positive-frequency solution, and multiplication by $-i$ for a negative-frequency solution.

To see the implications for what we mean by positive- and negative-energy 1-particle states, let us use the canonical second quantization and suppose that the free evolution of the quantum field is generated by the free evolution on the 1-particle subspace as in equation (5), which by the foregoing can be identified with the space $V_J$. In that case, with respect to $J$, the field evolves as the canonical second quantization $\Gamma$ of the unitary evolution:

$$f \rightarrow \exp(-JHt/\hbar)f$$

where $f \epsilon V_J$. The requirement that the energy be positive means that $H$ must be a positive operator. (It is self-adjoint because of Stone's theorem.) Equivalently, $H$ can have only positive (generalized) eigenfunctions. Since we know that the solution manifold $V$ contains positive- and negative-frequency solutions, this appears inconsistent with the infinitesimal form of the evolution (the Schrödinger equation). However, it is the complex structure $J$ which must now be used; that is, we now have:

$$Hf = J\hbar \frac{\partial f}{\partial t}. \tag{6}$$

When $J$ is given by multiplication by $-i$ on the negative frequency solutions,

of the form $f^- = w \exp(iEt/h)$, we obtain $Hf^- = Ef^-$. If $J$ has the natural action (multiplication by i) on the positive frequency states, then $H$ will be a positive operator on $V_J$; since (6) must yield the same evolution as the field equation, it is obvious that $H = -iJH_N$, where $H_N$ is the usual Dirac Hamiltonian. (The meaning of this notation will shortly become clear.)

To make further progress, we need a little more of the theory of complex structures on orthogonal and symplectic spaces. For the bosonic field, we assume that the solution manifold $V$ is a complex linear vector space equipped with a symplectic form $\omega$, with the usual complex structure given by multiplication by $i$ (the natural complex structure). In favourable cases one can find a canonical mapping $J$ on $V$ (i.e. with $\omega(Jf, Jg) = \omega(f, g)$) such that $J^2 = -\mathbb{I}$. It then automatically follows that

$$(f, g)_J = \omega(f, Jg)$$

is a symmetric form on $V$, and that:

$$\langle f, g \rangle_J = (f, g)_J + i\omega(f, g)$$

is sesquilinear on $V_J$; that is, it is sesquilinear with respect to multiplication of elements of $V$ by 'the complex numbers' $a + Jb, a, b \epsilon \mathbb{R}$. When $(.,.)_J$ is non-degenerate, it follows that $\langle .,. \rangle_J$ provides a Hilbertian norm, and we may complete to obtain a Hilbert space.

This is how we obtain the bosonic field theory; to obtain the fermionic theory we have instead of a symplectic form a *symmetric non-degenerate bilinear form*. For the Dirac field it is:

$$S(\psi, \phi) = \tfrac{1}{2}(\int \bar{\psi} \phi \mathrm{d}^3 x + \int \psi \bar{\phi} \,\mathrm{d}^3 x) \tag{7}$$

(here and in the following the spinor summation is suppressed); now $S(J\psi, \phi)$ is automatically antisymmetric and we may write:

$$\langle f, g \rangle_J = S(\psi, \phi) + iS(J\psi, \phi) \tag{8}$$

so that in both cases what fixes the Hilbert space and the properties of the operators is the complex structure $J$. In particular, we must choose $J$ such that the Hamiltonian $H$ is a positive operator.

This is at the 1-particle level; at the level of the Fock space, the creation and annihilation operators also depend critically on the complex structure $J$. To see this, we recall equation (1):

$$a(if) = -ia(f)$$
$$a^*(if) = ia^*(f).$$

Therefore in terms of the real-linear self-adjoint field:

$$\Phi(f) = (\hbar/2)^{1/2}[a(f) + a^*(f)].$$

We have[32]

$$a(f) = (2\hbar)^{-1/2}[\Phi(f) + i\Phi(if)]$$

$$a^*(f) = (2\hbar)^{-1/2}[\Phi(f) - i\Phi(if)] \tag{9}$$

and, as follows from the action of *a, a\**, we see that:

$$[a(f), a^*(g)]_\pm = \langle f, g \rangle ,$$

so that for fermions

$$[\Phi(f), \Phi(g)]_+ = \hbar S(f, g)$$

while for bosons

$$[\Phi(f), \Phi(g)]_- = i\hbar \omega(f, g).$$

In the present approach it is the field $\Phi$ that is considered fundamental, defined by an algebra independent of the complex structure *J*. If, in the foregoing, we consider the Hilbert space complex numbers given by *J*, we obtain from (9) new creation and annihilation operators:

$$a_J(f) = (2\hbar)^{-1/2}[\Phi(f) + i\Phi(Jf)]$$

$$a_J^*(f) = (2\hbar)^{-1/2}[\Phi(f) - i\Phi(Jf)] . \tag{10}$$

The field $\Phi$ is called the *Segal field.* It is real-linear, causal, and self-adjoint.[33] Assuming that it is given, the freedom in the particle interpretation corresponds to the freedom in the choice of *J*. Once *J* is chosen, then (10) tells us what are to count as creation and annihilation operators, and it is *J* that tells us the (anti)commutation relationships for $a_J$, because it defines the inner product $\langle .,. \rangle_J$ (via (8)). In this way, we find:[34]

$$\left[a_J(f), a_J^*(g)\right]_\pm = \langle f, g \rangle_J v$$

$$[a_N(f), a_N^*(g)]_\pm = \langle f, g \rangle_N . \tag{11}$$

As we have seen, if we choose the complex structure

$$J = iP^+ - iP^-$$

(where $P^\pm$ are projection operators on to the positive- and negative-frequency subspaces of $V_J$), the negative-frequency solutions no longer have negative energy in the sense of Schrödinger, i.e. according to (6). We shall call this choice of $J$ the *particle* complex structure. *We make the natural assumption that the negative-frequency states are the positive-energy antiparticle states.*

---

[32] Usually the factor $\sqrt{\hbar}$ appears only in the relationship between the Segal field and the creation and annihilation operators in the bosonic case. The issue here is a little subtle and I shall not pursue it here. For a conservative critique, see Rosenfeld (1963: 355).

[33] In non-relativistic quantum mechanics, the Segal field is of the form *P+Q*, where *P* and *Q* are the momentum and position operators. Its physical interpretation is obscure; mathematically, it is the generator of the Weyl algebra of the fields (boson case), and algebraically, it generates the Clifford algebra of the fields (fermion case).

[34] Quantities defined with respect to the natural (or particle) complex structure have subscript *N* (respectively, *J*).

Now consider the relationship between the creation and annihilation operators defined by the natural and particle complex structures (what we shall call the natural and particle creation and annihilation operators). The two are linked through the Segal field $\Phi$, which is independent of the complex structure. We see that:

$$(2\hbar)^{1/2} a_N(f) = \Phi(f) + i\Phi(if) = \Phi(f) + i\Phi[J(P^+ - P^-)f]$$

$$= \Phi(f^+) + \Phi(f^-) + i\Phi(Jf^+) - i\Phi(Jf^-)$$

$$= (2\hbar)^{1/2}[a_J(f^+) + a_J^*(f^-)]$$

and similarly for $a_N^*(f)$; that is,[35]

$$a_N(f) = a_J(f^+) + a_J^*(f^-)$$

$$a_N^*(f) = a_J(f^-) + a_J^*(f^+)\,.$$

We see that the natural annihilation and creation operators are combinations of annihilation and creation operators for particles and antiparticles; this is just the particle interpretation of the (usual) physical fields, for example the Dirac field $\psi$ and its adjoint $\psi^*$. We can therefore identify $a_N$ with $\psi$, etc., and $a_J(f^+)$ with $b$, $a_J(f^-)$ with $d$, and $a_J^*(f^-)$, $a_J^*(f^+)$ with $b^*$ and $d^*$, respectively.[36]

Now that we have the particle creation and annihilation operators, we can canonically second-quantize any 1-particle operator. Since for the identity $\mathbb{I}$,

$$\langle f^+, \mathbb{I}g^- \rangle_J = \langle f^+, g^- \rangle_J = 0\,,$$

we have from (5) that:

$$d\Gamma_J(\mathbb{I}) = \sum_k a_J^*(f_k) a_J(f_k) = \sum_k a_J^*(f_k^+) a_J(f_k^+) + \sum_k a_J^*(f_k^-) a_J(f_k^-).$$

Therefore with the obvious identifications:

$$d\Gamma_J(\mathbb{I}) = N^+ + N^-\,. \qquad\qquad .$$

Similarly, one sees that the total energy is positive: if $H_N$ is the Hamiltonian defined by the natural complex structure, then $-iJH_N$ is the particle Hamiltonian which is clearly positive (what we denoted $H$ in (6)), and its second quantization is the sum of the energy of all the particles and the energy of all the antiparticles. The 1-particle *charge operator,* on the other hand, is given by $iJe;$ obviously, the positive-frequency states have eigenvalue $-e,$ and the negative frequency states have eigenvalue $+e;$ the total charge operator is its canonical second quantization:

$$d\Gamma_J(iJe) = -eN^+ + eN^- .$$

---

[35] So far as I know, these equations first appeared in Bongaarts (1972). Segal treated only the real scalar field; for the treatment of the complex scalar field, see Saunders (1989, sect. 3.4).

[36] The identification is figurative (see fn. 17).

In fact, all the global kinematic observables can now be obtained by a canonical second quantization with respect to the particle complex structure. At the same time, since we now see that the physical fields are the natural annihilation and creation operators (with their particle interpretation fixed by the particle complex structure), the conventional theory can now be understood as a *canonical second quantization with respect to the natural complex structure followed by normal-ordering.* Both the Fock space action of the natural creation and annihilation operators and the normal-ordering process are defined by the particle complex structure.

It is helpful to prove this equivalence in detail. For this we need the explicit relationship between $\langle .,. \rangle_J$ and $\langle .,. \rangle_N$ . From (7), (8) it follows that

$$\langle f,g \rangle_N = \int \bar{f} \, g \, \mathrm{d}^3 x ,$$

whereas

$$\langle f,g \rangle_J = \langle f^+, g^+ \rangle_N + \langle g^-, f^- \rangle_N \qquad (12)$$

(note carefully the order of $f$ and $g$). We also need the relationship between 1-particle operators defined by the two complex structures. To determine this we consider only those observables $X$ which can be obtained from the classical theory and which preserve the two complex structures; $X$ then generates a one-parameter group of orthogonal transformations on $V$, which is the same as the group of transformations generated by $X_N$ on $V_N$ and by $X_J$ on $V_J$ (each as generators of unitary transformations with respect to the relevant complex structure). It then follows that

$$X_N = -iJX_J. \qquad (13)$$

(We have already used this relationship above.)

The equivalence in question is therefore between

$$: \mathrm{d}\Gamma_N(X_N) := \; : \sum_{k,j} a_N^*(f_k)\langle f_k, X_N f_j \rangle_N a_N(f_j):$$

(where we must normal-order with respect to the action of $a_N$ on $\mathcal{F}(V_J)$, that is in terms of the particle complex structure $J$), and

$$\mathrm{d}\Gamma_J(X_J) = \sum_{k,l} a_J^*(f_k)\langle f_k, X_J f_l \rangle_J a_J(f_l) .$$

In evaluating the first expression, the requirement that the Hamiltonian flow generated by $X$ preserves the complex structures (in particular $J$) means that $X_N$, $X_J$ do *not* connect negative- and positive- frequency states (i.e. $\langle f^+, X_N f^- \rangle_N = \langle f^+, X_J f^- \rangle_J = 0$). We then obtain:

$$: \mathrm{d}\Gamma_N(X_N) := \sum_{k,l} \big[ a_J^*(f_k^+) a_J(f_l^+) \langle f_k^+, X_N f_l^+ \rangle_N$$
$$- a_J^*(f_l^-) a_J(f_k^-) \langle f_k^-, X_N f_l^- \rangle_N \big]$$

(note carefully the order of the indices $k$, 1; the minus sign is due to normal ordering). From (12) and (13), it now follows that:

$$\langle f_k^+, X_N f_l^+ \rangle_N = \langle f_k^+, X_J f_l^+ \rangle_J$$
$$\langle f_k^-, X_N f_l^- \rangle_N = -\langle f_l^-, X_J f_k^- \rangle_J$$

and we obtain:

$$: \mathrm{d}\Gamma_N(X_N) := \sum_{k,l} \big[ a_J^*(f_k^+) a_J(f_l^+) \langle f_k^+, X_J f_l^+ \rangle_J$$
$$+ a_J^*(f_l^-) a_J(f_k^-) \langle f_l^-, X_J f_k^- \rangle_J \big] .$$

That is,

$$: \mathrm{d}\Gamma_N(X_N) := \mathrm{d}\Gamma_J(X_J) \tag{14}$$

as claimed.

# 8 Interpretation

What is the upshot of all this? We cannot say that the conventional theory is equivalent in all respects to the canonical second quantized theory with respect to the particle complex structure; this is true only for a limited class of global operators (which preserve particle number). In particular, the equivalence does not hold for local multiplicative operators, for these connect positive- and negative- frequency states. (They are 'odd' operators, in the sense of Schrödinger; equivalently, they do not commute with $J$.[37]) For these the RHS of (14), if considered a perturbation, would induce transitions from particle to antiparticle states, which would be a complete disaster. In this situation we must read back from the standard formalism; it seems that the complex structure must also change under the evolution. One must abandon or extend the canonical theory, because one does not have a fixed Hilbert

---

[37] There are intimate connections with the problem of defining a (position space) Born interpretation in relativistic theory; in the Foldy-Wouthuysen representation, equivalently the Newton-Wigner representation, $J$ is a local operator. To keep the discussion within reasonable bounds I shall not pursue the matter here; relevant material may be found in Segal (1964), Goodman and Segal (1965), Streater (1988), and Saunders (1989).

space (of the form $\mathcal{F}(V_J)$) to host a unitary evolution.

In the standard formalism, the LHS of (14) still makes sense as an operator on $\mathcal{F}(V_J)$: that is why it is possible develop a formal perturbation theory. From the canonical point of view, the standard formalism is a quantum mechanics over *two* complex structures; the natural complex structure is used to determine the creation and annihilation operators (which are the physical fields) and the canonically second quantized operators, but their action on Fock space is expressed in terms of the particle complex structure. (The physical Fock space is $\mathcal{F}(V_J)$, and not $\mathcal{F}(V_N) = \mathcal{F}(V)$.) Just because the two complex structures do not coincide, the operators $:d\Gamma_N(X_N):$ (or $d\Gamma_N(X_N)$ for that matter) contain products of linear combinations of particle creation and annihilation operators. Therefore, they describe pair creation and annihilation processes.

In the non-relativistic field theory, the particle complex structure is identical to the natural complex structure. In this case the physical fields are simply the canonical particle creation and annihilation operators; in the 1-particle theory, every real-linear operator extends to a complex-linear (or antilinear) operator. The implication of the foregoing is that these are special features of the non-relativistic limit which do not survive in the relativistic theory, not even in kinematics. There is also the implication that in some sense the relativistic theory is the canonical theory where the concept of *charge* replaces the concept of *particle*. What the natural creation and annihilation operators (the physical fields) create and destroy are *units of charge*. The natural complex structure attaches to the charge, and the particle complex structure to the particle number. This is borne out by the role of the total charge and number operators as generators of phase transformations (gauge transformations of the first kind). The charge generates phase transformations in the physical fields; i.e.

$$\exp(id\Gamma_N(\mathbb{1})\theta): a_N(f) \longrightarrow a_N(\exp(i\theta)f) = \exp(-i\theta)a_N(f)$$

(using the antilinearity of $a_N$ on $\mathcal{F}(V_N)$)), but the number operator $d\Gamma_J(\mathbb{1})$ generates phase transformations in the complex numbers that are used in the Hilbert space:

$$\exp(Jd\Gamma_J(\mathbb{1})\theta): a_J(f) \longrightarrow a_J(\exp(J\theta)f) = \exp(-i\theta)a_J(f)$$

(using the antilinearity of $a_J$ on $\mathcal{F}(V_J)$). For example, if we express the action of $\exp(id\Gamma_N(\mathbb{1})\theta)$ on $a_N$, in terms of the physical Hilbert space $\mathcal{F}(V_J)$, we find the curious properties that we noted were required in the Weinberg construction:

$$a_N(f) \longrightarrow a_N[\exp(i\theta)f] = a_J[\exp(J\theta f^+)] + a_J^*[\exp(-J\theta f^-)]$$
$$= \exp(-i\theta)a_J(f^+) + \exp(-i\theta)a_J^*(f^-).$$

Therefore, despite the fact that according to the canonical theory the annihilation and creation operators transform oppositely, the particle annihilation operator and antiparticle creation operator transform in the same sense under the gauge transformations of the physical fields. We see explicitly that the gauge transformations of the Dirac fields *are* induced by gauge transformations on the Hilbert space, but by the natural complex structure and not by the particle one. The conserved quantity is the charge;[38] the conserved quantity under rotations in the particle complex structure is the particle number. *Because the interaction Hamiltonian is required to be gauge-invariant with respect to the natural complex structure (i.e. bilinear in the natural annihilation and creation operators, the Dirac field, and its adjoint), it is the former and not the latter that is conserved in the dynamics.*

It is now easy to understand why the charge conjugation operator has such different properties in the 1-particle theory and in the standard formalism. Recall that this operator is unitary in the field theory, but antilinear in 1-particle theory; further, it does not (as it should) change the sign of the 1-particle charge. We can formulate these differences as follows. Whereas

$$\mathfrak{C} :: \mathrm{d}\Gamma_N(-e\mathbb{1}): \to -: \mathrm{d}\Gamma_N(-e\mathbb{1}):\,,$$

in the 1-particle theory the operator $-e\mathbb{1}$ is invariant under any unitary (or antiunitary) mapping defined on $V_N$. But there is a corresponding map on $V_J$, because $:\mathrm{d}\Gamma_N(-e\mathbb{1}): = \mathrm{d}\Gamma_J(ieJ)$ and the latter is a canonical second quantization (with no normal ordering); hence in particular $\Gamma(U)\mathrm{d}\Gamma_J(X)\Gamma(U)^{-1} = \mathrm{d}\Gamma(UXU^{-1})$ (where $U$ is unitary if and only if $\Gamma(U)$ is unitary). On $V_J$ the charge conjugation must be unitary.

This is just what we find; a simple calculation shows that $(Jf)^c = Jf^c$.[39] With respect to our new notion of complex numbers, the charge conjugation is linear, despite the complex conjugation contained in its action (intuitively, $\left(J(f_1^+ + f_2^-)\right)^c = -if_1^{+c} + if_2^{-c} \simeq -if_1^- + if_2^+ = J(f_1^- + f_2^+) = Jf^c)$. At the same time, the 1-particle charge now changes sign, because it is no longer the identity on $V_J$, but rather the operator $eiJ$, so that under $\mathcal{C}$ (using its linearity with respect to $J$, and antilinearity with respect to $N$),

$$\mathcal{C}eiJ\mathcal{C}^{-1} = -eiJ\,.$$

In summary, we are led to a point of view in which the (linear) standard formalism appears as a quantum mechanics of charge ('charge dynamics'), represented (through the normal ordering) in the kinematic limit as a quantum mechanics of particles. In both cases the quantum mechanics is in canonical second quantized form. On this view there is no non-trivial particle dynamics on a fixed complex-linear space (not even in Fock space[40]); that is, there is no particle Hilbert space description of the dynamics. In the kinematic case, only observables that commute with $J$ can be defined.

---

[38] Strictly speaking, we should be second-quantizing $-e\mathbb{1}$ rather than $\mathbb{1}$ to obtain the physical charge.

[39] In detail: in a representation with $\gamma^0$ real, one has $\gamma^0 = \gamma^{0t}$, $\gamma^0\gamma^{\mu\dagger}\gamma^0 = \gamma^\mu$ ($\dagger$ is complex conjugation followed by matrix transposition). Recalling that $\mathcal{C}\gamma^\mu\mathcal{C}^{-1} = -\gamma^{\mu t}$, $P^\pm = \pm i\gamma^\mu p_\mu/m$, then $(Jf)^c = \mathcal{C}\gamma^{0t}\overline{Jf} = -i\mathcal{C}\gamma^{0t}\left(\gamma^\mu p_\mu\right)^{\dagger t}\gamma^{0t}\mathcal{C}^{-1}f^c = i\gamma^\mu p_\mu f^c = Jf^c$.

To pursue these implications would take us too far afield and into difficult terrain. For what it is worth, C*-algebra theory also leads to a similar conclusion, in so far as one finds that the evolution cannot be unitarily implemented within any one representation, and in semi- classical quantum gravity the metric dependence of the particle interpretation of quantum fields has a natural expression in terms of inequivalent complex structures on Hilbert space.[41] The almost complete failure of the constructive programme in quantum field theory - in which the existence of a particle Fock space is an axiom - is in itself remarkable.[42] The present theory offers a new perspective which is both simple and radical. It is simple because (for linear fields) the standard formalism can be understood as a canonical second quantization; contact with the 1-particle theory is preserved. Relativistic quantum theory is cast into a form almost identical to the non-relativistic theory. It is radical because the modification of the complex structure ramifies throughout the interpretive and mathematical framework of the theory.[43] It may be that one can formulate a theory in which the complex structure is changing with time; however, we must first formulate an interpretation of a Hilbert space theory in which the usual local operators are no longer complex-linear. If these things can be done, we will have an interpretation of relativistic theory of some beauty; the complex numbers of the Hilbert space theory will determine the particle interpretation and will themselves change with time, subject to the dynamics. In this way the existence of particles is built into the dynamics. For linear interactions this could even be studied at the 1-particle level.

This is a radical view; a more conservative approach is that there is no exact theory of interactions, not because complex linearity fails but because the standard formalism is incomplete.[44] Actually, the best one can do is claim that, *so long as one has a scattering situation,* there is a relativistic quantum theory on a particle Hilbert space. There is, after all, a clear enough intuition of a particle dynamics, in which an incoming particle state evolves into a coherent superposition of outgoing particle states, all within a fixed Hilbert space. This seems to conform to the basic framework of elementary quantum mechanics.

---

[40] In the present framework Fock space is defined as $\mathcal{F}(V_J)$; $J$ cannot be preserved by the evolution.

[41] See Ashtekar and Magnon (1975), Woodhouse (1980: 284-7). For a self-contained introduction along more conventional lines, see Birrel and Davies (1982).

[42] This programme has culminated in the result that $\lambda\phi^4$ theory in $3 + 1$ dimensions, a super-renormalizable quantum field theory, has in fact only the trivial solution S =1. A similar conclusion seems to hold for QED itself. For a review see e.g. Huang (1989).

[43] It seems to me that the absence of complex linearity during interactions, or with respect to certain operators, *may* have a decisive bearing on measurement theory. As I have argued elsewhere (Saunders 1988), the measurement problem requires exact mathematics; we know that the world is not Galilean, and that relativistic effects, however small, are always present in the measurement process. See also the conjecture of Maxwell (1988), according to which scattering processes differing in particle number for the outgoing states should *not* be considered coherent, and that of Penrose (1989), where the '1-graviton' criterion (for longitudinal gravitons) likewise signals the breaking of coherence. The present framework may provide a theoretical basis for these conjectures.

But charge superselection, the fact that only gauge-invariant operators count as physical observables, is not an incidental feature to this picture; on the contrary, charge superselection, and with it the construction of a field theory over the tensor product of the particle and antiparticle Fock spaces, can be looked on as a way of rewriting quantum mechanics over a 'number' field with imaginary unit of the form $J$, in the conventional format of the non-relativistic theory. The unrestricted complex linearity of the non-relativistic theory is preserved through elaborate constraints on what is to count as an observable. The distinction between the two Fock spaces still depends on the decomposition of the field into positive- and negative-frequency parts (this is why we need a scattering situation); each Fock space has the natural complex structure, but the usual self-adjoint operators of elementary theory cannot act on these spaces. (They map states out of each Fock space altogether.)

Is this *unrestricted* complex linearity? This may better be considered a fiction; if it is so qualified at the relativistic level, the unrestricted version that appears when we descend to the non-relativistic limit may be an idealization of no fundamental physical significance.

I shall conclude with a last look at the negative-energy sea. The complex structure $J$ leads to a reinterpretation of the negative-energy solutions, much as did the hole theory; consider now their relationship. Here we must bear in mind the fact that the Dirac negative-energy sea enforces a dynamical interpretation as well as a kinematic one. The properties of antimatter were deduced both from the assumption of the Dirac vacuum and from dynamical considerations (transitions to vacant negative-energy states with the emission of energy). By means of the exclusion principle, these properties could be deduced from the physical picture of the vacuum.

Consider now the following theorem, which holds when $V$ is a *finite* (say s)-dimensional space. (I owe this observation to Professor G. Segal of the Mathematical Institute, University of Oxford.) In the fermion case the natural Fock space is the exterior algebra over $V$, i.e. $V_0 \bigoplus_{n=1}^{s} \Lambda_{i=1}^{n} V_i$ (where each $V_i$ is a copy of $V$ and $V_0$ is the one-dimensional vector space $\mathbb{C}$).

---

[44] This point of view is particularly natural in path-integral quantization. Although the present theory indicates that the standard formalism actually describes the dynamics in the natural Fock space ('charge dynamics'), the path-history space appears to preserve complex linearity at the expense of unidirectional evolution in time.

Consider each $n$-particle subspace $\Lambda_{i=1}^{n} V_i$; there is a unique *antilinear* isomorphism with the $(s-n)$-particle subspace, or a linear isomorphism $\approx$ with the complex conjugate $(s-n)$-particle subspace:

$$\bigwedge_{i=1}^{n} \overline{V}_i \approx \bigwedge_{i=1}^{s-n} V_i \ .$$

(15)

(This is the generalization of the Hodge •-operator to the complex case.[45]) For the negative-frequency subspace of $V$ (the antiparticle states), this is equivalent to replacing the action of multiplication by $i$ by multiplication by $-i$, so it also implements the change from the natural complex structure on these states to the particle complex structure. It does so, however, by replacing a vector in the n-antiparticle state by a vector in the $(s-n)$-particle state. But this is the prescription of the Dirac hole theory: the vacuum state, with $n = 0$ and $s$ infinite, is replaced by the $(s-n = \infty)$-particle state[46]- the negative-energy sea. We can guess what is going on; even when $n \neq 0$, (15) continues to ensure that the 'Dirac dual' $(s-n)$-particle state,[47] using the natural complex structure, is equivalent to the physical n-antiparticle state using the particle complex structure. The holes behave as do the antiparticles.[48]

We see that the Dirac vacuum enforces a different notion of complex numbers at the Hilbert space level from that suggested by the non-relativistic theory. We may conclude that the negative-energy sea is what the particle vacuum looks like using the wrong notion of complex numbers (the natural complex structure). If the particle vacuum is to appear really empty, then we must use the particle complex structure at the Hilbert space level.

---

[45] The Hodge • transformation is an isomorphism resting on the canonical duality between differential forms and tangent vectors. For real spaces this duality is bilinear; in the complex case it is sesquilinear. (There is no positive-definite non-singular bilinear form on $V \times V$, with $V$ complex.)

[46] Note that, for $s$ finite, both states are rays, i.e. one-dimensional.

[47] I.e. the state with $(s-n)$ negative-frequency electrons present, where s is the total number of states available; alternatively, the Dirac vacuum with $n$ holes.

[48] This conclusion appears valid even when $J$ is changing with time, that is for arbitrary evolutions. At each instant there is a correspondence between the holes and the antiparticles.